\documentclass[3p,times,preprint]{elsarticle}

\usepackage{fullpage,amssymb,amsmath,amsthm}
\usepackage{amsfonts,amssymb,mathtools}
\usepackage{fullpage,amssymb,amsmath,amsthm}

\usepackage{color,svgcolor}
\usepackage[plainpages=true,bookmarks=false]{hyperref}
\makeatletter
\hypersetup{
pdftitle={\@title},
breaklinks=true,
colorlinks=true,
linkcolor=darkred,
citecolor=darkblue,
linkcolor= darkred,
anchorcolor= green,
filecolor= green,
urlcolor= darkblue,
menucolor= green
}
\makeatother

\usepackage[capitalize]{cleveref}

\makeatletter
\newcommand*{\rem}{%
  \nonscript\mskip-\medmuskip\mkern5mu%
  \mathbin{\operator@font rem}\penalty900\mkern5mu%
  \nonscript\mskip-\medmuskip
}
\makeatother

\usepackage{aliascnt}
\newtheorem{theorem}{Theorem}
\numberwithin{theorem}{section}
\newaliascnt{lemma}{theorem}\newtheorem{lemma}[lemma]{Lemma}\aliascntresetthe{lemma}
\newaliascnt{proposition}{theorem}\aliascntresetthe{proposition}
\newaliascnt{corollary}{theorem}\aliascntresetthe{corollary}
\newaliascnt{example}{theorem}\aliascntresetthe{example}

\crefname{lemma}{Lemma}{Lemmas}
\crefname{proposition}{Proposition}{Propositions}
\crefname{corollary}{Corollary}{Corollaries}
\crefname{example}{Example}{Examples}
\theoremstyle{definition}
\newaliascnt{algo}{theorem}\newtheorem{algo}[algo]{Algorithm}\aliascntresetthe{algo}
\theoremstyle{remark}
\newaliascnt{remark}{theorem}\aliascntresetthe{remark}
\theoremstyle{plain}
\crefname{algo}{Algorithm}{Algorithms}
\crefname{remark}{Remark}{Remarks}
\crefname{enumi}{Step}{Steps}
\crefname{enumii}{Step}{Steps}
\newcommand{\HankelSolve}{\hyperref[alg:hankelsolve]{HankelSolve}}
\newcommand{\Extension}{\hyperref[alg:extension]{Extension}}
\newcommand{\Transition}{\hyperref[alg:transition]{Transition}}
\newcommand{\softO}{\tilde{O}}

\newcommand{\Tr}{\mathrm{T}}
\DeclareMathOperator{\diag}{diag}
\newcommand*{\doi}[1]{\href{https://doi.org/#1}{doi:\nolinkurl{#1}}}

\newcommand\R{{\sf R}}
\newcommand\K{{\sf K}}

\newcount\hour\newcount\minutes\newcount\temp
\hour=\time \divide\hour by 60 \temp=\hour \multiply\temp by 60
\minutes=\time \advance\minutes by-\temp

\begin{document}

\begin{frontmatter}

\title{On computing the determinant over a ring}

\author{Gilles Villard}

\affiliation{{CNRS, ENS de Lyon, Inria, U. Claude Bernard Lyon 1, LIP UMR 5668, 69342, Lyon cedex 07, France}}

\begin{abstract}
An algorithm is presented for computing the determinant and the characteristic polynomial of a matrix $A$ over an arbitrary 
commutative ring $\R$. The algorithm uses $O(n^\omega)$ ring additions, subtractions and multiplications, where $\omega>2$ is a
feasible exponent for matrix multiplication, that is, such that two $n\times n$ matrices over~$\R$ can be
multiplied using $O(n^{\omega})$ ring operations.
The algorithm involves no division and no zero test, and its control depends only on~$n$: for each~$n$
it is a straight-line program.

The approach combines a polynomial matrix point of view and a block Krylov point of view on the characteristic
polynomial, which makes Strassen's elimination of divisions efficient.
A bivariate matrix $\lambda I-L(t)$, obtained from $A$ by a homotopy, is compressed in $\log n$ steps, each of
which halves the dimension while only doubling the degree in~$\lambda$.
The cost of the homotopy is controlled by truncating the power series in~$t$
at an order that only grows with the compression steps, from a constant to~$n$.
\end{abstract}

\end{frontmatter}

\section{Introduction} \label{sec:intro}

The determinant of an $n \times n$ matrix over a field can be computed using $O(n^\omega)$ arithmetic operations~\cite{BuHo74},
where~$\omega>2$ is a feasible exponent for matrix multiplication. That is, two $n\times n$ matrices
can be multiplied using $O(n^{\omega})$ operations in the field, or in the ring of the entries in the
division-free setting considered below.
The same bound holds for the characteristic polynomial~\cite{NeigerPernet21}.
The algorithms achieving these bounds, Gaussian elimination for the determinant~\cite{BuHo74} and the polynomial
matrix approach of~\cite{NeigerPernet21} for the characteristic polynomial (see also~\cite{KeGe85,PeSt07}),
use divisions and zero tests, for choosing nonzero pivots and, more generally, for computing rank profiles.

Over a commutative ring, where divisions are not available, the determinant and the characteristic polynomial
can be computed using $\softO(n^{\omega+1})$ ring operations, where $\softO$ hides logarithmic factors.
This bound is obtained in two different ways, by the method of Samuelson-Berkowitz~\cite{berkowitz1984} 
(see also~\cite{chistov1985}), and by Strassen's technique for eliminating
divisions~\cite{St73}. The latter applies directly to the determinant, which is a polynomial of degree~$n$ in the entries,
and costs a factor about~$n$: every operation becomes an operation on power series
truncated at order~$n+1$. With fast polynomial multiplication over an arbitrary ring~\cite{CaKa91},
this gives $\softO(n^{\omega+1})$ ring operations~\cite{Kal92}.

Combinatorial approaches give another point of view on division-free computation. Such an interpretation of
Berkowitz's algorithm is given in~\cite{Val92}, and an algorithm based on clow
sequences is proposed in~\cite{MaVi97soda,MaVi99} (see also~\cite{Rote01}). 

The determinant over an arbitrary commutative ring has algebraic branching
programs of polynomial size~\cite{Tod92}, \cite{Val92}. The direct
construction of~\cite{MaVi97soda} has recently been improved in~\cite{Ike25}.
These approaches focus on a different model 
and do not seem to lead to an improvement of the $\softO(n^{\omega+1})$ bound.

The gap between the field and the ring case was partly closed by Kaltofen~\cite{Kal92}, who applies
Strassen's technique to Wiedemann's algorithm~\cite{Wie86} with a baby steps/giant steps strategy, and obtains the
characteristic polynomial using $\softO(n^{\omega/2+2})$ ring operations (see also~\cite {Seifullin2003}).
With block projections,  the cost is further reduced to $O(n^{\eta})$~\cite{KaVi05},
where $\eta$ depends on~$\omega$ and on the exponent of rectangular matrix multiplication.
With the exponents known at the time, this gave $\eta\approx 2.7$, that is, $\eta-\omega\approx 0.32$.
For a fixed exponent of rectangular matrix multiplication, this difference $\eta-\omega$ increases when
$\omega$ decreases, and it is~$1/2$ for $\omega=2$, whatever the rectangular exponent~\cite[Table~6.1]{KaVi05}.

Whether the exponent~$\omega$ can be reached without divisions is a question that goes back to
Berkowitz~\cite[Section~4]{berkowitz1984}. He conjectured more: that the characteristic polynomial over an
arbitrary ring can be computed by circuits of size $O(n^{\omega})$ (up to logarithmic factors) and depth
$O(\log^2 n)$. We do not address the depth here.

Kaltofen and Storjohann list the division-free computation
of the characteristic polynomial in~$n^{\omega+o(1)}$ operations as an open problem~\cite{KaSt15}.
The question is natural since the determinant and the coefficients of the characteristic
polynomial, are polynomials with integer coefficients in the entries of the matrix: only ring operations
are involved in their definition. The  point therefore is whether divisions 
bring an intrinsic advantage.

\subsection{Our result}
\label{subsec:intro_result}

We work over a commutative ring~$\R$ with unit.
We analyse our algorithms by counting the number of operations from $\R$
on an algebraic random access machine. The only operations 
are additions, subtractions and multiplications, with no division and no zero test. The control
depends only on the dimension~$n$, so that for each~$n$ the algorithm is a straight-line program~\cite[Section~4.1]{BCS97}.

In this model, we show that divisions do not essentially help for computing the determinant. 

For  $A\in\R^{n\times n}$,
\cref{alg:charpoly} computes the characteristic polynomial of~$A$, hence its determinant, using
$O(n^{\omega})$ ring operations (\cref{thm:charpoly}), the matrix multiplication algorithm being used as
a black box (\cref{subsec:prelimin}). For $\omega=2$, the same
analysis gives $O(n^2\log^3 n\log\log n)$ operations.
The control of \cref{alg:charpoly} depends only on~$n$. This is not the case of the known algorithms,
even for the determinant alone over a field: the $O(n^{\omega})$ triangular factorization of Bunch and Hopcroft~\cite{BuHo74} relies on pivoting,
as does~\cite{NeigerPernet21} for the characteristic polynomial.

The algorithm is division-free and computes the determinant as a polynomial in the entries of~$A$,
hence the derivative inequality of Baur and Strassen~\cite{BaSt83} applies. The adjoint of~$A$,
whose entries are, up to transposition, the partial derivatives of the determinant, is thus also
computed using $O(n^{\omega})$ ring operations.

Our approach reconciles two points of view. 

On one side, the polynomial matrix approach to the determinant~\cite{GJV03,NeigerPernet21} starts from $\lambda I-A$, an $n\times n$ matrix
of degree one, and follows a divide-and-double principle. In $O(\log n)$ steps, the dimension is halved
while the degree only doubles, so that the amount of data stays $O(n^2)$ and each step costs a few
matrix products. The same principle is used, e.g., in high-order lifting~\cite{Sto03},
for computing kernel bases~\cite{ZLS12} or the Hermite normal form~\cite{LVZ17}. It is also close to
the doubling of the Krylov chains in Keller-Gehrig's algorithm~\cite{KeGe85}. These algorithms
use divisions and zero tests, and we do not see how to apply a homotopy
\`a la Strassen to them without losing a factor about~$n$.

On the other side, Coppersmith's block Wiedemann approach~\cite{Coppersmith94}, in
its division-free version by~\cite{KaVi05}, works with the block Krylov sequence $\{X^{\mathsf T}A^kY\}_{k\geq 0}$
of a pair of projections $X,Y\in\R^{n\times m}$, from which the characteristic polynomial is recovered as the
determinant of an $m\times m$ matrix generating polynomial of degree about $n/m$ (\cref{subsec:intro_coppersmith}). The projections are chosen once and for all, with a fixed block size~$m$. Divisions are avoided by
Strassen's technique: $A$ is replaced by a matrix $L(t)$ of degree one in~$t$ with $L(0)=M$, a fixed
matrix for which all the elements inverted by the algorithm with divisions are known units of~$\R$.
The computations are then done on power series in~$t$, where these elements remain
invertible, and the result is read on $\det(\lambda I-L(t))$, a polynomial of degree at
most~$n$ in~$t$. The terms of the sequence are then matrices over $\R[t]$, and the generating polynomial
is an $m\times m$ matrix polynomial in~$\lambda$ with coefficients in the power series ring $\R[[t]]$
(over a field~$\K$, it would be in $\K(t)[\lambda]^{m\times m}$, with denominators that are units of $\K[[t]]$).
With this homotopy, the $2n/m$ terms of the sequence are computed by baby steps and giant steps on power
series truncated at order~$n+1$.

We combine the two points of view, with two ideas (\cref{subsec:overview}). The first idea is to run the scheme of~\cite{KaVi05}
at $\log n$ successive levels, with block sizes $m=n,n/2,\ldots,1$, {\em without recomputing the block Krylov
sequence at each level}: the projections of a level are made of columns of those of the previous level,
hence the sequence needed at level $m/2$ is extracted from the generating polynomial computed
at level~$m$, by expanding the corresponding rational matrix series. This replaces the baby steps and
giant steps, uses only matrix polynomial operations, and halves the block size while doubling the
degree, as in the divide-and-double principle. {\em The second idea concerns the homotopy variable~$t$: the terms of the sequence that are used are kept exact,
while the generating polynomial is truncated}. The terms of the sequence at level~$m$, with $d=n/m$, are polynomials in~$t$ of
degree less than~$4d$, and the generating polynomial, whose coefficients are power series in~$t$, is only
needed modulo $t^{4d}$: all the computations at level~$m$ are truncated at order~$4d$ instead of~$n+1$.
The amount of data at each level is $O(n^2)$, and the cost of a level decreases geometrically with~$m$, up to
logarithmic factors, when $\omega>2$, and sums to $O(n^{\omega})$.

The homotopy and the projections are modified with respect to~\cite{KaVi05}, so that Strassen's technique
applies at every level (\cref{sec:realizations}). {\em The choice of sparse structured projections $X$ and~$Y$, whose nonzero entries are interleaved with a
cyclic structure of~$M$}, is central in our approach, both for the efficiency  and for the elimination of divisions.

\subsection{Coppersmith's block Wiedemann approach}
\label{subsec:intro_coppersmith}

We give the main ingredients needed in the rest of the paper. More details and references can be
found in~\cite[Section~2]{KaVi05} and~\cite[Section~5]{NSSV24}. We present them over a field~$\K$.
Over a ring, the properties we need will follow from our choice of the homotopy and of the projections
(\cref{sec:mingen,sec:realizations}).
Let $L\in\K^{n\times n}$, let $n=md$, and let $X,Y\in\K^{n\times m}$ be two projections. Coppersmith's block Wiedemann approach
 relies on the sequence of $m\times m$ matrices
\begin{equation}\label{eq:Hk}
H_k=X^{\Tr}L^kY,\qquad k\ge 0,
\end{equation}
and on its matrix generating polynomials. A matrix polynomial
$D(\lambda)=\sum_{j=0}^{d}D_j\lambda^j\in\K[\lambda]^{m\times m}$ is a (right) matrix generating polynomial of
the sequence, in short a generating polynomial, if
\begin{equation}\label{eq:generator}
\sum_{j=0}^{d}H_{k+j}D_j=0\qquad\text{for all } k\ge0 .
\end{equation}
Let $\K((\lambda^{-1}))$ be the field of Laurent series in $\lambda^{-1}$, in which
\begin{equation}\label{eq:laurent}
H(\lambda)=X^{\Tr}(\lambda I-L)^{-1}Y=\sum_{k\ge0}H_k\,\lambda^{-k-1}.
\end{equation}
The coefficient of $\lambda^{-k-1}$ in $H(\lambda)D(\lambda)$ is $\sum_{j=0}^{d}H_{k+j}D_j$,
hence $D$ is a generating polynomial if and only if the product $H(\lambda)D(\lambda)$ is a polynomial matrix,
of degree less than~$d$. When $D$ is nonsingular, this is a right matrix fraction description
\begin{equation}\label{eq:fraction}
X^{\Tr}(\lambda I-L)^{-1}Y=\sum_{k\ge0}X^{\Tr}L^kY\,\lambda^{-k-1}=N(\lambda)\,D(\lambda)^{-1},\qquad N\in\K[\lambda]^{m\times m},\ \deg N<d.
\end{equation}
If the $n\times n$ block Hankel matrix $\mathcal H=(H_{i+j})_{0\le i,j<d}$ is invertible, then there is a
unique monic generating polynomial of degree~$d$, obtained by solving a linear system with
matrix~$\mathcal H$, and it satisfies
\begin{equation}\label{eq:detD}
\det D(\lambda)=\det(\lambda I-L).
\end{equation}

The reader may refer to~\cite[Section~6.5, p.\,439]{Kailath80}
for the underlying realization theory.
The rank properties of block Hankel matrices and the generating polynomials
they determine are studied in~\cite[Section~2]{KaVi05}, and in~\cite[Section~5]{NSSV24} where generating
polynomials are used as matrices of relations.
For the sake of completeness, and since we work over a ring, \eqref{eq:detD} is proven in \cref{sec:mingen}.

Equation~\eqref{eq:detD} displays the compression mechanism behind the divide-and-double principle:
the $n\times n$ matrix $\lambda I-L$ of degree one is replaced by the $m\times m$ matrix~$D$ of degree~$d=n/m$,
with the same determinant.
Equations~\eqref{eq:fraction} and~\eqref{eq:detD} illustrate precisely how we combine the two points of view:
the middle term of~\eqref{eq:fraction} is the sequence of the block Krylov approach, while
the right-hand side is a matrix fraction description as in the polynomial matrix approach. The
characteristic polynomial is reached through the denominator~$D$ by~\eqref{eq:detD}. The generating
polynomial is obtained from the first $2d$ terms of the sequence by a matrix version of the Euclidean
algorithm in~\cite[Section~3]{KaVi05}, whereas the polynomial matrix approach relies on approximant
bases in~\cite{GJV03} (see also~\cite{BeLa94}), and on kernel bases~\cite{ZLS12} in~\cite{NeigerPernet21}.
In our case, the truncated generating polynomial is computed by solving a block Hankel linear system by lifting
(\cref{alg:hankelsolve}).

\subsection{Overview of the algorithm}
\label{subsec:overview}

Let $A\in\R^{n\times n}$. Following Strassen's technique and~\cite[Section~2]{KaVi05}, we do not work with $A$ directly but with the
homotopy
\begin{equation}\label{eq:homotopy}
L(t)=M-tAM=(I-tA)M\in\R[t]^{n\times n},\qquad L(0)=M,
\end{equation}
where $M\in\R^{n\times n}$ is chosen such that $\det M=(-1)^{n-1}$ and so as not to cause divisions
(see \hyperref[it:homotopy]{\em (i)} below).
\Cref{alg:charpoly} computes the bivariate polynomial
$$\det(\lambda I-L(t))\in\R[\lambda,t],$$ 
which is monic of degree~$n$ in~$\lambda$ and of degree at most~$n$
in~$t$. Its constant coefficient in~$\lambda$ is $(-1)^n\det L(t)=-\det(I-tA)$,
and the characteristic polynomial of~$A$ is the reversal $\lambda^n\det(I-A/\lambda)$ of the latter.

We assume $n$ is a power of $2$, otherwise we replace $A$ by $A'=\diag(A,I_{n'-n})$ with $n'$ the next power of $2$, $n\le n'<2n$.
The determinant is unchanged, $\mathrm{adj}\,A'=\diag(\mathrm{adj}\,A,\ \det A\cdot I)$,
and $\det(I-tA')=\det(I-tA)(1-t)^{n'-n}$. The characteristic polynomial is thus recovered after
a multiplication by the unit series $(1-t)^{-(n'-n)}$ modulo $t^{n+1}$. The padding
multiplies the cost of the algorithm by at most a constant factor.

Starting from $L(t)$, the algorithm is organized around a sequence of $\log_2 n+1$ matrix polynomials
$D^{(m)}$ in~$\lambda$, of dimensions $m=n,n/2,\ldots,1$ and degrees $d=n/m$ in~$\lambda$, whose coefficients are
power series in~$t$ computed modulo suitable powers of~$t$ (see \hyperref[it:truncation]{\em (ii)}), and
whose determinants are congruent to $\det(\lambda I-L(t))$ modulo these powers. They follow the divide-and-double principle of
\cref{subsec:intro_result}. From one level to the next, the dimension is divided by two while the degree
in~$\lambda$ is multiplied by two.
These successive compressions are closely related to those of the determinant algorithm
of~\cite[Section~4.2.2]{GJV03} (obtained there with divisions).

At each level, $D^{(m)}$ is, modulo the truncation in~$t$, the monic generating polynomial of degree~$d$,
 of the sequence $H^{(m)}_k=X_m^{\Tr}L^kY_m$, in the sense of~\eqref{eq:generator}.  
It is determined by the first $2d$ terms of the sequence (\cref{lem:gen2,lem:lifting}).

For the transition from level $m$ to level $m/2$ (\cref{step:loop} in \cref{alg:charpoly}),
three main ingredients are needed to obtain the target complexity bound without divisions.
From the first $2d$ terms of the sequence at level~$m$,
\cref{step:gen} computes $D^{(m)}$ modulo $t^{\sigma_m}$ ($\sigma _m$ is defined in \hyperref[it:truncation]{\em (ii)}).
The latter is then used in \cref{step:extend} to extend the sequence to its first $4d$ terms.
In \cref{step:extract}, the sequence at level~$m/2$ is finally extracted from the extended one at level~$m$.
In short, the algorithm is as follows.

\begin{algo}{\em Characteristic polynomial} \label{alg:charpoly}\hfill\\
{\em Input:} $A\in\R^{n\times n}$,  $n$ a power of two (up to a padding)\\{\em Output:} $\det(\lambda I-A)$
\begin{enumerate}
\item\label{step:init} $L\leftarrow M-tAM$, at level $m=n$ ($d=1$), $H^{(n)}_0=I_n$ and $H^{(n)}_1=L$
\item\label{step:loop} For $m=n,n/2,\ldots,2$, with $d=n/m$: 
\hfill{\em $\triangleright$ \Transition$(m)$, \cref{alg:transition}}\\[-0.34cm]
  \begin{enumerate}
  \item\label{step:gen} compute $D^{(m)}$ modulo $t^{4d}$ from $H^{(m)}_0,\ldots,H^{(m)}_{2d-1}$ 
  \hfill{\em $\triangleright$ \hyperref[alg:hankelsolve]{lifting} in~$t$ for the generating polynomial} \\[-0.34cm]
  \item\label{step:extend} compute $H^{(m)}_{2d},\ldots,H^{(m)}_{4d-1}$ from $D^{(m)}$ 
  \hfill{\em $\triangleright$ \hyperref[alg:extension]{extension}, Newton iteration in~$\lambda$} \\[-0.22cm]
  \item\label{step:extract} extract $H^{(m/2)}_0,\ldots,H^{(m/2)}_{4d-1}$ from $H^{(m)}_0,\ldots,H^{(m)}_{4d-1}$ 
  \hfill{\em $\triangleright$ extraction, no arithmetic cost}
  \end{enumerate}
\item\label{step:last} At level $m=1$ ($d=n$), compute $D^{(1)}$ modulo $t^{n+1}$ from $H^{(1)}_0,\ldots,H^{(1)}_{2n-1}$ \\
$c(t)\leftarrow -D^{(1)}_0=\det(I-tA)$, return $\lambda^n c(1/\lambda)$.
\end{enumerate}
\end{algo}
We now detail the three ingredients mentioned above.

\begin{description}
\item \phantomsection\label{it:homotopy}{\em (i) Homotopy and projections (\cref{sec:realizations}).}
We take $L(t)=M - tAM$, where $M$ is the $n\times n$ cyclic permutation matrix. 
This homotopy is the one in~\cite[Sec.\,6]{KaVi05}, with a different $M$. 
The projections $X_m$ and $Y_m$ are made of unit vectors at well chosen positions, whose interleaving  
allows for an efficient transition from level $m$ to level $m/2$. 
With this choice the block Hankel matrix of the sequence is a permutation matrix at $t=0$, at every level (\cref{sec:realizations}). 
The minimal generating polynomial is then the solution of a linear system whose constant term is invertible,
which is solved by lifting in~$t$, without divisions.
\item \phantomsection\label{it:truncation}{\em (ii) Truncation at the right orders (\cref{sec:transition}).}
The degrees in~$\lambda$ are controlled by the dimension--degree trade-off explained above. 
We also need to control the sizes in~$t$. To do this, 
we keep the terms $X_m^{\Tr}L^kY_m$ of the sequences exact from one level to the next,
while the generating polynomials, whose coefficients are power series in~$t$, are truncated to an appropriate order~$\sigma_m$. 
Exactness is needed since each level requires twice the order of the previous one.
It is possible since the terms satisfy $\deg_t(X_m^{\Tr}L^kY_m)\le k$.
At level $m$, the generating polynomial $D^{(m)}$ is used to extend the sequence to its first~$4d$ terms,
which is what the next level needs. At level $m/2$ the degree of $D^{(m/2)}$ in $\lambda$ is indeed $2d$, and it is determined by $2\cdot 2d=4d$ terms (at level $m=1$, with $d=n$, the characteristic polynomial is reconstructed from $2n=4 \cdot n/2$ terms provided by level~2).
Since these terms are polynomials of degree less than $4d$ in~$t$, it is 
sufficient to work with a generating polynomial $D^{(m)}$ truncated at order $\sigma _m = 4d$ only.  

The truncation order in $t$ and the degree in $\lambda$ grow with $d$ only.
A level $m$ first consists of a lifting process, to compute the truncated generating polynomial, followed by a
Newton iteration to extend the sequence. 
Both are divide-and-conquer processes whose aggregate work, summed over their $O(\log d)$ recursion
depths, is equivalent to that of $O(\log d)$ products of $m\times m$ matrices whose entries are bivariate
polynomials of degree~$O(d)$ both in~$\lambda$ and in~$t$, that is, with $O(d^2)$ coefficients.
The cost is thus $m^{\omega}d^2$, up to factors logarithmic in~$d$.
The total cost is dominated by the first level $m=n$, where $d=1$, which gives $O(n^{\omega})$ when $\omega>2$. 
\item \phantomsection\label{it:extraction}{\em (iii) New sequence by extraction (\cref{lem:extract}).}
The generating polynomial at level $m$ allows us to extend the sequence from $2d$ to $4d$ terms.
The matrix $L$ is the same at every level. The nested projections $X_m$ and $Y_m$ (see \hyperref[it:homotopy]{\em (i)}) are chosen in order to  
ensure that the sequence terms at level $m/2$ are obtained by simply selecting rows and columns of the extended sequence at level $m$,
at no arithmetic cost.
Hence $A$ is used only at the first level, and the sequences are never recomputed from~$L$.
\end{description}

\subsection{Outline}
\label{subsec:outline}

In \cref{sec:mingen} we take a closer look at \cref{eq:fraction} and \cref{eq:detD}  over a ring. We relate the generating polynomials of a sequence to its block Hankel matrix, and give the proofs
we need for Coppersmith's block Wiedemann approach.
The homotopy and the projections are defined in \cref{sec:realizations}, where we show that they satisfy
the conditions necessary for the efficient and division-free computation.
\Cref{sec:transition} is devoted to the algorithm for transition from level~$m$ to level~$m/2$: the lifting in~$t$
for the generating polynomial, the extension of the sequence by a Newton iteration in~$\lambda$, and their costs.
The main theorem is proven in \cref{sec:mainthm}, by $\log n$ applications of the transition, followed at the
last level, where $m=1$, by the solution by lifting of a Hankel system with $1\times1$ blocks, which gives $\det(\lambda I-L(t))$, and by
the reconstruction of the characteristic polynomial from its constant coefficient in~$\lambda$.

\subsection{Preliminaries}
\label{subsec:prelimin}

To analyse the algorithms we count operations in the ring~$\R$ on an algebraic random access machine: additions, subtractions and
multiplications in~$\R$ have unit cost, the constants $0$ and $1$ are available. 
The straight-line program model would be just as relevant, since the algorithm has no branching on ring
elements.

We assume a fixed algorithm that multiplies two $n\times n$ matrices over~$\R$ using $O(n^{\omega})$ ring
operations, for a feasible exponent $\omega>2$, and we use it as a black box (see also
\cite[Section~1.1]{NeigerPernet21}). The case $\omega=2$ is considered after \cref{thm:charpoly}.

Let $\mathsf M(h)$ be such that two polynomials of degree less than~$h$ over any $\R$-algebra can be
multiplied using $O(\mathsf M(h))$ operations on their coefficients: additions, multiplications of two
coefficients, and multiplications by integers. Cantor and Kaltofen's algorithm gives
$\mathsf M(h)=h\log h\log\log h$~\cite{CaKa91}. We assume that $\mathsf M(h)/h$ is
nondecreasing and $\mathsf M(ch)\le c^2\mathsf M(h)$ for integers $c\ge1$~\cite[Section~8.3, Eq.\,(8)]{GaGe99}.
Applied to polynomials whose coefficients are $m\times m$ matrices over~$\R$, with the black box for the
coefficient products, this multiplies two $m\times m$ matrices of polynomials of degree less than~$h$ using
$O(m^{\omega}\mathsf M(h))$ operations in~$\R$. The finer count of~\cite{CaKa91}, $O(h\log h)$ coefficient
products only, is not needed here: the bound $O(n^{\omega})$ of \cref{thm:charpoly} is insensitive to these
logarithmic factors, and holds for any $\mathsf M(h)=O(h^{\omega/2-\varepsilon})$ with $\varepsilon>0$.

If the matrix multiplication algorithm is bilinear, it is valid over any commutative
$\R$-algebra and can be run directly over a univariate polynomial ring over~$\R$. The same bound then holds
with any polynomial multiplication algorithm over~$\R$ satisfying the above hypotheses on~$\mathsf M$, since
only products of polynomials with coefficients in~$\R$ are needed.

By Kronecker substitution, bivariate polynomials of degree less than $\delta$ in~$\lambda$ modulo
$t^{\sigma}$ are handled with $h=O(\delta\sigma)$.

We write $I_k$ for the $k\times k$ identity matrix, or simply~$I$ when the dimension is clear from the
context, and $X^{\Tr}$ for the transpose of a matrix~$X$. Logarithms are in base~$2$.

\bigskip
\noindent
{\bf Acknowledgments.}
The author would like to thank Vincent Neiger and Clément Pernet for the very enriching discussions on the Krylov approach and polynomial matrices.

\smallskip
\noindent
{\bf Use of AI tools.}
During the preparation of this manuscript, the author used Claude (Anthropic),
in several versions during 2026.  
The tool was first used in an attempt to compare structured matrix methods for computing the generating polynomial.
This did not lead to a choice of $M$, $X$, $Y$ allowing one to keep the approach of~\cite{KaVi05} with a matrix
Euclidean algorithm, and the author's initial option, presented here, was retained.
It was useful for bibliographical searches, for shortening \cref{sec:mingen} (whose
ingredients are quite classical) and the proof of \cref{lem:Hzero}, and for the complexity analysis and its assumptions. 
It was used in the final phase for drafting the text from the author's instructions, in a presumably better English than
the author's, sentence by sentence. Word by word, symbol by symbol, all AI-generated English suggestions were reviewed
and, if applicable, modified by the author before deciding whether to include them in the manuscript.

\smallskip
\noindent
{\bf Note.}
On November 3, 2026, shortly after the author had shared an early draft of this manuscript with a few
colleagues, Mark Giesbrecht, who had not seen it, informed him that on October 30, 2026 he had
independently prompted ChatGPT Astra on this problem and obtained an algorithm of similar complexity, with a
Lean proof. Despite apparent similarities, the two constructions differ in essential ways. A main difference
is the block Hankel solver: we lift from a constant inverse, while the other algorithm refines inverse
generators by Newton iteration. The homotopies also differ: we use $M-tAM$ (\cite{KaVi05}, up to the choice
of~$M$), while the other construction (called deformation) adds a nilpotent matrix, $N+tA$. This changes how the characteristic
polynomial is read off at the end, from the constant coefficient and a reversal here~\cite{KaVi05}, from the
homogeneous component of degree~$n$ there. We apply the general identity $\det D=\det(\lambda I-L)$
(\cref{lem:detD}), while the other construction uses the Cayley--Hamilton theorem at the last level.
Finally, we obtain $O(n^{\omega})$ where the other algorithm is stated with $\softO(n^{\omega})$. The
logarithmic factors are probably not intrinsic to it.

\section{Generating polynomials and characteristic polynomial}
\label{sec:mingen}

Let $\mathcal S$ be a commutative ring, $L\in\mathcal S^{n\times n}$, and $X,Y\in\mathcal S^{n\times m}$ with $n=md$.
In this section we give conditions under which a matrix $D$ as in~\eqref{eq:fraction}
and~\eqref{eq:detD} exists, is unique once normalized, and can be computed from the first $2d$ terms
of the sequence $X^{\Tr}L^kY$ in~\eqref{eq:laurent}. 
This is well understood over a field, see for instance \cite[Chapter~6]{Kailath80} for matrix fraction descriptions
and their link with the controllability and observability matrices,
and \cite[Section~2]{KaVi05}, \cite[Sections~5 and 7]{NSSV24} and references therein for matrix sequences. 
We recall the main properties for the sake of completeness. 
The results of this section hold for an arbitrary commutative ring~$\mathcal S$ and arbitrary $L$, $X$, $Y$.
They are specialized in \cref{sec:realizations}, where $\mathcal S$ is $\R[[t]]$, or $\R[t]/(t^{\sigma})$ when
working modulo $t^{\sigma}$.

Let $H_k=X^{\Tr}L^kY\in\mathcal S^{m\times m}$ for $k\ge0$.
The observability and controllability Krylov matrices are
\[
\mathcal O=\begin{bmatrix} X^{\Tr}\\ X^{\Tr}L\\ \vdots\\ X^{\Tr}L^{d-1}\end{bmatrix}\in\mathcal S^{n\times n},
\qquad
\mathcal C=\begin{bmatrix} Y & LY & \cdots & L^{d-1}Y\end{bmatrix}\in\mathcal S^{n\times n},
\]
and their product is the block Hankel matrix
\begin{equation}\label{eq:HOK}
\mathcal H=\mathcal O\,\mathcal C=
\begin{bmatrix}
H_0     & H_1    & \cdots & H_{d-1}\\
H_1     & H_2    & \cdots & H_d\\
\vdots  &        &        & \vdots\\
H_{d-1} & H_d    & \cdots & H_{2d-2}
\end{bmatrix}\in\mathcal S^{n\times n}.
\end{equation}
In the terminology of linear systems \cite[Chapter~6]{Kailath80}, $(L,Y,X^{\Tr})$ is a realization
of the transfer matrix $X^{\Tr}(\lambda I-L)^{-1}Y$, and $\mathcal C$ and $\mathcal O$ are its controllability
and observability matrices (when $\mathcal H=\mathcal O\,\mathcal C$ is invertible, the realization is
controllable and observable).
We also let $\bar H=(H_d,\ldots,H_{2d-1})^{\Tr}\in\mathcal S^{n\times m}$.

Over~$\mathcal S$, \eqref{eq:laurent} and~\eqref{eq:fraction} are identities in $\mathcal S((\lambda^{-1}))^{m\times m}$,
where $\mathcal S((\lambda^{-1}))$ is the ring of Laurent series in $\lambda^{-1}$ with coefficients in~$\mathcal S$. 
The matrix $\lambda I-L$ is invertible over this ring,
with inverse $\sum_{k\ge0}L^k\lambda^{-k-1}$, and so is any monic matrix polynomial $D$ of degree~$d$.

\begin{lemma}\label{lem:gen2}
Assume that $\mathcal H$ is invertible over $\mathcal S$. Then $\mathcal O$ and $\mathcal C$ are invertible,
and the system $\mathcal H\,\bar D=-\bar H$ has a unique solution $\bar D=(D_0,\ldots,D_{d-1})^{\Tr}\in\mathcal S^{n\times m}$.
Setting $D_d=I_m$, we obtain
\begin{equation}\label{eq:LYD}
\sum_{j=0}^{d}L^jYD_j=0,
\end{equation}
hence $D(\lambda)=I_m\lambda^d+\sum_{j=0}^{d-1}D_j\lambda^j$ is a monic generating polynomial of degree~$d$ for the sequence $\{H_k\}_{k\geq 0}$, 
i.e.~\eqref{eq:generator} holds, and it is the unique one.
The equality between matrix fractions in~\eqref{eq:fraction} follows.
\end{lemma}
\begin{proof}
Since $\det\mathcal H=\det\mathcal O\det\mathcal C$ is a unit of $\mathcal S$, 
$\det\mathcal O$ and $\det\mathcal C$ are units, and $\mathcal O$, $\mathcal C$ are invertible.
The solution $\bar D=-\mathcal H^{-1}\bar H$ is unique.
Conversely, any monic generating polynomial of degree~$d$ satisfies~\eqref{eq:generator} for $k=0,\ldots,d-1$,
which is the system $\mathcal H\,\bar D=-\bar H$. This gives the uniqueness of~$D$.
The $i$-th block row of $\mathcal H\,\bar D+\bar H$ is
$\sum_{j=0}^{d-1}H_{i-1+j}D_j+H_{i-1+d}D_d=X^{\Tr}L^{i-1}v$, where $v=\sum_{j=0}^{d}L^jYD_j$, 
hence $\mathcal H\,\bar D+\bar H=\mathcal O\,v$.
Since $\mathcal H\,\bar D+\bar H=0$ and $\mathcal O$ is invertible, $v=0$, which is~\eqref{eq:LYD}.
Multiplying $v=0$ on the left by $X^{\Tr}L^k$ gives $\sum_{j=0}^{d}H_{k+j}D_j=0$ for all $k\ge0$, and~\eqref{eq:generator}
is obtained.  The equality between matrix fractions in~\eqref{eq:fraction} then follows from the equivalence stated after~\eqref{eq:generator}.
\end{proof}

Note that, over a field, the generating polynomial of \cref{lem:gen2} is also minimal in the sense of~\cite[Section~2.1]{KaVi05}. 
Indeed, the determinant of a basis of the module of right vector generators has degree equal to the rank of
the infinite block Hankel matrix~\cite[Eq.~(2.6)]{KaVi05},
and $\deg\det D=md=n$. Minimality is however not needed in what follows.

\begin{lemma}\label{lem:detD}
Assume that $\mathcal H$ is invertible over $\mathcal S$, and let $D$ be a generating polynomial as in \cref{lem:gen2}.
Then $\det D(\lambda)=\det(\lambda I-L)$, which is~\eqref{eq:detD}, and in particular $\det D_0=(-1)^n\det L$.
\end{lemma}
\begin{proof}
By~\eqref{eq:LYD}, $L^dY=-\sum_{j=0}^{d-1}L^jYD_j$, hence $L\,\mathcal C=\mathcal C\,\Gamma$ with the block companion matrix 
\[
\Gamma=\begin{bmatrix}
0 &        &   & -D_0\\
I_m & \ddots &   & -D_1\\
  & \ddots & 0 & \vdots\\
  &        & I_m & -D_{d-1}
\end{bmatrix}\in\mathcal S^{n\times n}.
\]
Since $\mathcal C$ is invertible (\cref{lem:gen2}), $\lambda I-L$ and $\lambda I-\Gamma$ are similar,
and $\det(\lambda I-L)=\det(\lambda I-\Gamma)$. 
The block companion matrix $\lambda I-\Gamma$ is a linearization of $D(\lambda)$,
see \cite[Section~6.3.4, p.\,393]{Kailath80}, whose proof is valid over any commutative ring. 
The matrix $\lambda I-\Gamma$ is equivalent, by unimodular matrices, to $\diag(I_m,\ldots,I_m,D(\lambda))$,
hence $\det(\lambda I-\Gamma)=u\det D(\lambda)$ for a unit $u$ of $\mathcal S[\lambda]$. Both determinants are
monic of degree~$n$ in~$\lambda$, and $\det D$ is monic, so that $u\det D$ has degree $n+\deg u$. Therefore, 
$\deg u=0$ and $u=1$ and this gives $\det(\lambda I-\Gamma)=\det D(\lambda)$. 
Taking $\lambda=0$ gives $\det D_0=(-1)^n\det L$.

\end{proof}

\section{The homotopy and the projections}
\label{sec:realizations}

In this section, $\mathcal S$ is $\R[[t]]$ or $\R[t]/(t^{\sigma})$ for some $\sigma\ge1$. We define the
homotopy $L(t)$ and the projections $X_m$, $Y_m$, which are chosen so as to satisfy the conditions of
\cref{sec:mingen} at every level~$m$. The block Hankel matrix $\mathcal H$ is invertible over~$\mathcal S$.
with a value at $t=0$ that is a permutation matrix.
A matrix over~$\mathcal S$ is invertible if and only if its value at $t=0$ (that is, its constant coefficient in~$t$) is invertible over~$\R$.

Let $e_1,\ldots,e_n$ be the canonical basis of $\mathbb Z^n$, with indices taken modulo~$n$
and represented in $\{1,\ldots,n\}$. We have $e_{j+n}=e_j$ and in particular $e_0=e_n$.
Let $M\in\mathbb Z^{n\times n}$ be the cyclic permutation matrix defined by $Me_j=e_{j+1}$, $1\le j\le n$,
it is the companion matrix of $\lambda^n-1$.
Our homotopy is
\[
L(t)=M-tAM=(I-tA)M\in\R[t]^{n\times n}, ~~~L(0)=M.
\]
For $n=md$, the projections at level~$m$ are
\[
X_m^{\Tr}=\begin{bmatrix}
u_d^{\Tr} &        &   \\
          & \ddots &   \\
          &        & u_d^{\Tr}
\end{bmatrix}\in\mathbb Z^{m\times n},
\quad \!\!\!\!
M=\begin{bmatrix}
0 &        &        & 1\\
1 & 0      &        &  \\
  & \ddots & \ddots &  \\
  &        & 1      & 0
\end{bmatrix}\in\mathbb Z^{n\times n},
\quad \!\!\!\!
Y_m=\begin{bmatrix}
u_1 &        &   \\
    & \ddots &   \\
    &        & u_1
\end{bmatrix}\in\mathbb Z^{n\times m},
\]
where $u_1$ and $u_d$ are the first and last canonical vectors of $\mathbb Z^{d}$.
We have 
$X_m=[\,e_d,\ e_{2d},\ \ldots,\ e_{md}\,]$ and $Y_m=[\,e_1,\ e_{d+1},\ \ldots,\ e_{(m-1)d+1}\,]$. 
At level~$m$, we apply \cref{sec:mingen} with $X=X_m$ and $Y=Y_m$.

The corresponding $H_k=X_m^{\Tr}L^kY_m$, block Hankel matrix~$\mathcal H$ and generating polynomial~$D$
depend on~$m$. As we did for $D^{(m)}$ in the introduction, we may write $H_k^{(m)}$ and $D^{(m)}$,
but we omit the superscript when the level is clear from the context.

The homotopy is the one of \cite[Section~6]{KaVi05} for the characteristic polynomial, with a different matrix~$M$.
The latter and the projections $X_m$, $Y_m$ are chosen so as to ensure, at every level~$m$,
the properties needed by the algorithm.

\begin{description}
\item -
The sequence terms $X_m^{\Tr}L^kY_m$ at level $m/2$ are extracted from those at level~$m$
by selecting rows and columns, and they satisfy $\deg_t(X_m^{\Tr}L^kY_m)\le k$ (\cref{lem:extract}), which allows the truncation
at the right orders in~\cref{sec:transition}.
\item - The block Hankel matrix $\mathcal H$ is a permutation matrix at $t=0$, hence invertible over~$\mathcal S$ (\cref{lem:Hzero}). 
\Cref{lem:gen2,lem:detD} therefore apply, and the lifting  in~$t$ for computing the generating polynomial needs no division (\cref{alg:hankelsolve}).
\item - 
Moreover, since the generating polynomial $D$ is monic, its reversal $Q(\lambda)=\lambda^dD(1/\lambda)$ satisfies $Q(0)=I_m$,
which makes the extension of the sequences division-free (\cref{alg:extension}). Note that 
this property comes from the normalization $D_d=I_m$, whatever the choice of $M$, $X_m$ and $Y_m$ (\cref{lem:gen2}).
\end{description}

\begin{lemma}\label{lem:extract}
For all $k\ge0$, the entries of $X_m^{\Tr}L^kY_m$ are polynomials in~$t$ of degree at most~$k$.
If $m$ is even, then for all $k\ge0$ and $1\le i,j\le m/2$,
\[
\bigl(X_{m/2}^{\Tr}L^kY_{m/2}\bigr)_{i,j}=\bigl(X_m^{\Tr}L^kY_m\bigr)_{2i,\,2j-1}.
\]
\end{lemma}
\begin{proof}
The entries of $L(t)=M-tAM$ have degree at most one in~$t$, hence those of $L^k$ have degree at most~$k$.
For the second assertion, the columns of $X_{m/2}$ are the even columns of $X_m$, and the columns of $Y_{m/2}$ are the odd columns of~$Y_m$.
Indeed, at level $m/2$ the integer $d=n/m$ is replaced by $2d$. The $i$-th column of $X_{m/2}$ is $e_{i\cdot 2d}=e_{(2i)\cdot d}$, which 
is column $2i$ of~$X_m$.
The $j$-th column of $Y_{m/2}$ is $e_{(j-1)\cdot 2d+1}=e_{((2j-1)-1)d+1}$, which is column $2j-1$ of~$Y_m$.
\end{proof}

Let $J\in\mathbb Z^{n\times n}$ be the block reversal matrix, with $d\times d$ blocks of dimension $m\times m$,
whose blocks are $I_m$ on the anti-diagonal and $0$ elsewhere.

\begin{lemma}\label{lem:Hzero}
Let $H_k=X_m^{\Tr}L^kY_m$, and let $\mathcal H$ be the block Hankel matrix~\eqref{eq:HOK} for $X=X_m$, $Y=Y_m$. For $0\le k\le 2d-2$, one has $H_{d-1}(0)=I_m$, and $H_k(0)=0$ if $k\neq d-1$.
Consequently $\mathcal H(0)=J$ (with $J^{-1}=J$), and $\mathcal H$ is invertible over~$\mathcal S$.
\end{lemma}
\begin{proof}
Since $L(0)=M$, we have $H_k(0)=X_m^{\Tr}M^kY_m$, whose entry $(i,j)$ is $e_{id}^{\Tr}M^ke_{(j-1)d+1}=e_{id}^{\Tr}e_{(j-1)d+1+k}$.
It is nonzero, and then equal to one, if and only if $(j-1)d+1+k\equiv id \bmod n$ (with representation in $\{1,\ldots,n\}$).
This requires $k+1\equiv 0\bmod d$, which for $0\le k\le 2d-2$ means $k=d-1$. 
For $k=d-1$ the condition becomes $jd\equiv id\bmod md$, that is, $m$ divides $j-i$. 
Since $1\le i,j\le m$, the latter means $i=j$, hence $H_{d-1}(0)=I_m$ and $H_k(0)=0$ otherwise.
By~\eqref{eq:HOK}, $\mathcal H(0)$ has the blocks $H_{d-1}(0)=I_m$ on its anti-diagonal and zero blocks elsewhere, that is, $\mathcal H(0)=J$.
Since $J$ is invertible, so is $\mathcal H$ over~$\mathcal S$.
\end{proof}

\section{Transition from level $m$ to level $m/2$}
\label{sec:transition}

Let $m\ge 1$ be a power of two, $d=n/m$, and $H^{(m)}_k=X_m^{\Tr}L^kY_m$ for $k\ge0$.
For $m\ge2$, the transition (\cref{step:loop} in \cref{alg:charpoly}) takes as input the first~$2d$ terms
$H^{(m)}_0,\ldots,H^{(m)}_{2d-1}$, exact in~$t$, and returns the first $4d$ terms
$H^{(m/2)}_0,\ldots,H^{(m/2)}_{4d-1}$ of the sequence at level $m/2$, again exact. These are the terms
needed at level~$m/2$, where the generating polynomial has degree~$2d$.
It consists of three steps: the computation of $D^{(m)}$ modulo $t^{4d}$ by lifting in~$t$
(\cref{step:gen}), the extension of the sequence at level~$m$ to its first $4d$ terms by power series
expansion in~$\lambda$ (\cref{step:extend}), and the extraction (\cref{step:extract}).
Working modulo $t^{4d}$ is sufficient since the terms $H^{(m)}_k$ for $k<4d$, hence also the extracted
$H^{(m/2)}_k$, have degree less than~$4d$ in~$t$ (\cref{lem:extract}).

The lifting, \HankelSolve{} and \cref{lem:lifting} below, is valid for any level $m\ge1$. It is also used
at the last level $m=1$, with $d=n$, in \cref{step:last}.
We solve the linear system $\mathcal H\bar D=-\bar H$ of \cref{lem:gen2} at level~$m$ over
$\mathcal S=\R[t]/(t^{4d})$, which gives $D^{(m)}$ modulo $t^{4d}$. The reduction modulo $t^{4d}$ is a ring
homomorphism from $\R[[t]]$ onto~$\mathcal S$, and by \cref{lem:Hzero,lem:gen2} the system has a unique
solution over both rings, so that the solution over~$\mathcal S$ is the reduction of the one over~$\R[[t]]$.

\begin{algo}  {\em HankelSolve$(\sigma,B)$} (\cref{alg:charpoly}, \cref{step:gen})\label{alg:hankelsolve}\hfill\\
{\em Input:} $\sigma\ge1$, $B\in\R[t]^{n\times m}$, and $\mathcal H$ given by $H_0,\ldots,H_{2d-2}$, with $\mathcal H(0)=J$ \\
{\em Output:} $x\in\R[t]^{n\times m}$ such that $\mathcal Hx\equiv B\bmod t^{\sigma}$
\hfill{\em $\triangleright$ $D^{(m)} \bmod t^{\sigma}$ is deduced from \cref{lem:gen2}}
\begin{enumerate}
\item\label{step:hs-base} If $\sigma=1$, return $J\,B(0)$
\item $\sigma'\leftarrow\lceil\sigma/2\rceil$, $y\leftarrow$ \HankelSolve$(\sigma',B\bmod t^{\sigma'})$
\item\label{step:hs-res} $r\leftarrow(B-\mathcal Hy)\bmod t^{\sigma}$ \hfill{\em $\triangleright$ divisible by $t^{\sigma'}$}
\item $z\leftarrow$ \HankelSolve$(\sigma-\sigma',r/t^{\sigma'})$
\item return $y+t^{\sigma'}z$ 
\end{enumerate}
\end{algo}

\Cref{alg:hankelsolve} is the divide-and-conquer lifting of~\cite[Section~3]{BostanChyzakOllivierSalvySchostSedoglavic2007}
(see also~\cite{MoCa79} over a domain, and~\cite{Dix82} for the $p$-adic case), and is used for computing the generating polynomial.

\begin{lemma}\label{lem:lifting}
Let $H_k=X_m^{\Tr}L^kY_m$ for $0\le k\le 2d-1$, let $\mathcal H$ be as in \cref{lem:Hzero} and $\bar H=(H_d,\ldots,H_{2d-1})^{\Tr}$,
and let~$\sigma\ge1$.
The generating polynomial $D^{(m)}$ modulo $t^{\sigma}$ can be computed 
using $O(m^{\omega}\mathsf M(d\sigma)(1+\log\sigma))$ operations in~$\R$.
\end{lemma}
\begin{proof}
We first show that \HankelSolve$(\sigma,B)$ returns $x$ such that $\mathcal Hx\equiv B\bmod t^{\sigma}$.
Since $\mathcal H(0)=J=J^{-1}$ by \cref{lem:Hzero}, \cref{step:hs-base} is correct and requires no arithmetic operation.
Then $\mathcal H(y+t^{\sigma'}z)=B-r+t^{\sigma'}\mathcal Hz\equiv B\bmod t^{\sigma}$.
The solution is unique since $\mathcal H$ is invertible over $\R[t]/(t^{\sigma})$, $\bar D=$ \HankelSolve$(\sigma,-\bar H)$
and $D^{(m)} \bmod t^{\sigma}$ is deduced from \cref{lem:gen2}.

For the cost, let $h(\lambda)=\sum_{k=0}^{2d-2}H_k\lambda^k$ and, for $x=(x_0,\ldots,x_{d-1})^{\Tr}$,
$\tilde x(\lambda)=\sum_{j=0}^{d-1}x_j\lambda^{d-1-j}$. For $0\le i\le d-1$, the $(i+1)$-th block of $\mathcal Hx$, namely $\sum_{j=0}^{d-1}H_{i+j}x_j$,
is the coefficient of $\lambda^{d-1+i}$ in $h(\lambda)\tilde x(\lambda)$ (see for instance~\cite[Ch.\,2, Sec.\,5]{BiPa94},
only a middle product is needed).
A product by $\mathcal H$ modulo $t^{\sigma}$ thus reduces to the product of two polynomials of degree less than $2d$ in~$\lambda$
whose coefficients are $m\times m$ matrices over $\R[t]/(t^{\sigma})$, which costs
$O(m^{\omega}\mathsf M(d\sigma))$ (\cref{subsec:prelimin}).
The recursion has depth $\lceil\log_2\sigma\rceil\le1+\log\sigma$, and at each depth the precisions sum to~$\sigma$. 
Since $\mathsf M(h)/h$ is nondecreasing, 
each depth costs $O(m^{\omega}\mathsf M(d\sigma))$, which gives the bound.
\end{proof}

The next step in the transition is the extension of the sequence $\{H_k^{(m)}\}_{k\geq0}$ at level~$m$ from its first $2d$ terms
to its first $4d$ terms (\cref{step:extend}). We can now use the generating polynomial for that. 

The latter polynomial, $D^{(m)}$, is monic by \cref{lem:gen2}.
Let $Q(\lambda)=\lambda^dD^{(m)}(1/\lambda)=I_m+D_{d-1}\lambda+\cdots+D_0\lambda^d$ be the reversal of~$D^{(m)}$,
and $G(\lambda)=X_m^{\Tr}(I-\lambda L)^{-1}Y_m=\sum_{k\ge0}H^{(m)}_k\lambda^k=\lambda^{-1}H(1/\lambda)$, with $H$ as in~\eqref{eq:laurent}.
For $k\ge0$, the coefficient of $\lambda^{k+d}$ in $G(\lambda)Q(\lambda)$ is $\sum_{j=0}^{d}H_{k+j}D_j$,
which is zero by~\eqref{eq:generator}. Hence $G(\lambda)Q(\lambda)=P(\lambda)$ with $\deg_{\lambda}P<d$ (see \cite[Lemma~12.8]{GaGe99} for the scalar case),
and $P$ only depends on $H_0,\ldots,H_{d-1}$.
Since $D^{(m)}$ is monic  we have $Q(0)=I_m$. 
The matrix~$Q$ is thus invertible as a power series in~$\lambda$, and the expansion of $G=PQ^{-1}$ allows use to extend the sequence
(equivalently, the terms are given by the quotient in a division with remainder by the monic~$D^{(m)}$ \cite[Section~9.1]{GaGe99}). 

The inverse of $Q$ is computed by Newton iteration in~$\lambda$ \cite[Section~9.1]{GaGe99}. 
Note that \cref{alg:extension} recomputes the terms $H_0,\ldots,H_{2d-1}$, which are already known. 
This does not change the cost, asymptotically. 
Middle products could be used, both in the Newton iteration and for the last product~\cite{hanrot2004}, and 
would save constant factors.

\begin{algo}  {\em Extension$(\tau)$} (\cref{alg:charpoly}, \cref{step:extend})\label{alg:extension}\hfill\\
{\em Input:} $\tau\ge d$, $H_0,\ldots,H_{d-1}$, and $D^{(m)}$ modulo $t^{\sigma}$ \\
{\em Output:} $H_0,\ldots,H_{\tau-1}$ modulo $t^{\sigma}$
\begin{enumerate}
\item $Q\leftarrow\lambda^dD^{(m)}(1/\lambda)$, and $P\leftarrow\bigl(\sum_{k=0}^{d-1}H_k\lambda^k\bigr)\,Q\bmod\lambda^d$
\item\label{step:ext-newton} $R\leftarrow I_m$, for $\ell=1,2,4,\ldots$ while $\ell<\tau$: $R\leftarrow R\,(2I_m-QR)\bmod\lambda^{2\ell}$
\hfill{\em $\triangleright$ $R\equiv Q^{-1}\bmod\lambda^{\tau}$}
\item return the coefficients of $PR\bmod\lambda^{\tau}$
\end{enumerate}
\end{algo}

In \cref{alg:extension}, $D^{(m)}$ is only known modulo $t^{\sigma}$, and all the computations are carried out
over $\R[t]/(t^{\sigma})$. Since the reduction modulo $t^{\sigma}$ is a ring homomorphism, the identity
$G=PQ^{-1}$ holds over this ring with $D^{(m)}$ replaced by its reduction, hence the algorithm returns the
terms $H^{(m)}_k$ modulo $t^{\sigma}$, for any $\sigma\ge1$.
They are exact as soon as $\deg_tH_k<\sigma$ for $k<\tau$. In this section $\deg_tH_k\le k$ (\cref{lem:extract}),
and we use it with $\sigma\ge\tau$.

\begin{lemma}\label{lem:extension}
Let $\tau\ge d$ and $\sigma\ge1$. Given $H^{(m)}_0,\ldots,H^{(m)}_{d-1}$ and $D^{(m)} \bmod t^{\sigma}$,
\cref{alg:extension} computes $H^{(m)}_0,\ldots,H^{(m)}_{\tau-1}$ modulo $t^{\sigma}$
using $O(m^{\omega}\mathsf M(\tau\sigma))$ operations in~$\R$.
If $\deg_tH^{(m)}_k<\sigma$ for $k<\tau$, these terms are exact.
\end{lemma}
\begin{proof}
All computations are over $\mathcal S=\R[t]/(t^{\sigma})$, where~\eqref{eq:generator} still holds by reduction modulo $t^{\sigma}$.
\Cref{step:ext-newton} is the Newton inversion of \cite[Algorithm~9.3]{GaGe99}, written for matrix coefficients.
Since $Q(0)=I_m$, the initial value $R=I_m$ satisfies $I_m-QR\equiv0\bmod\lambda$.
If $I_m-QR\equiv0\bmod\lambda^{\ell}$, then $I_m-QR(2I_m-QR)=(I_m-QR)^2\equiv0\bmod\lambda^{2\ell}$,
hence \cref{step:ext-newton} gives $R\equiv Q^{-1}\bmod\lambda^{\tau}$, and $PR\equiv G\bmod\lambda^\tau$.
If $\deg_tH_k<\sigma$ for $k<\tau$, the terms computed modulo $t^{\sigma}$ are exact.
This holds for instance if $\sigma\ge\tau$, since $\deg_tH_k\le k$ (\cref{lem:extract}).

Each iteration of \cref{step:ext-newton} consists of two products of $m\times m$ matrices of polynomials
of degree less than $2\ell$ in~$\lambda$ over~$\mathcal S$, which cost $O(m^{\omega}\mathsf M(\ell\sigma))$ (\cref{subsec:prelimin}). Since $\mathsf M(h)/h$ is nondecreasing, 
the sum over $\ell$ is $O(m^{\omega}\mathsf M(\tau\sigma))$.
The products in the first and last steps have the same cost.
\end{proof}

The last step of the transition is the extraction of the sequence at level $m/2$ (\cref{step:extract}),
which requires no arithmetic operation by \cref{lem:extract}.

\begin{algo}  {\em Transition$(m)$} (\cref{alg:charpoly}, \cref{step:loop})\label{alg:transition}\hfill\\
{\em Input:} $m\ge2$, $d=n/m$, $H^{(m)}_0,\ldots,H^{(m)}_{2d-1}$ \\
{\em Output:}  $H^{(m/2)}_0,\ldots,H^{(m/2)}_{4d-1}$
\begin{enumerate}
\item\label{step:tr-gen} $\bar D\leftarrow$ \HankelSolve$(4d,-\bar H)$, which gives $D^{(m)}\bmod t^{4d}$
\hfill{\em $\triangleright$ \cref{lem:lifting}}
\item\label{step:tr-ext} $H^{(m)}_0,\ldots,H^{(m)}_{4d-1}\leftarrow$ \Extension$(4d)$, with $\sigma=4d$
\hfill{\em $\triangleright$ \cref{lem:extension}}
\item\label{step:tr-extract} $\bigl(H^{(m/2)}_k\bigr)_{i,j}\leftarrow\bigl(H^{(m)}_k\bigr)_{2i,2j-1}$ for $1\le i,j\le m/2$ and $0\le k\le 4d-1$
\hfill{\em $\triangleright$ \cref{lem:extract}}
\end{enumerate}
\end{algo}

\begin{lemma}\label{lem:transition}
Let $m\ge2$ be a power of two and $d=n/m$.
Given the terms $H^{(m)}_0,\ldots,H^{(m)}_{2d-1}$, \cref{alg:transition} computes the terms
$H^{(m/2)}_0,\ldots,H^{(m/2)}_{4d-1}$ using $O(m^{\omega}\mathsf M(d^2)(1+\log d))$ operations in~$\R$.
\end{lemma}
\begin{proof}
By \cref{lem:lifting} with $\sigma=4d$, \cref{step:tr-gen} gives $D^{(m)}$ modulo $t^{4d}$
using $O(m^{\omega}\mathsf M(4d^2)(1+\log(4d)))$ operations.
By \cref{lem:extension} with $\tau=\sigma=4d$, \cref{step:tr-ext} gives 
$H^{(m)}_0,\ldots,H^{(m)}_{4d-1}$ using $O(m^{\omega}\mathsf M(16d^2))$ operations.
\Cref{step:tr-extract} requires no arithmetic operation, and gives the exact terms at level~$m/2$ using \cref{lem:extract}.
The bound follows from $\mathsf M(ch)\in O(\mathsf M(h))$ for a constant~$c$, and from $1+\log(4d)=3+\log d\le3(1+\log d)$.
\end{proof}

\section{Characteristic polynomial}
\label{sec:mainthm}

Putting the pieces of previous sections together, we obtain \cref{alg:charpoly}. 
At the first level $m=n$ we have $d=1$ and $X_n=Y_n=I_n$, hence $H^{(n)}_0=I_n$ and $H^{(n)}_1=L$, which are exact.
At the last level $m=1$ we have $d=n$, and~$D^{(1)}=\det(\lambda I-L)$ by \cref{lem:detD}.
Since the entries of $\lambda I-L$ have total degree at most one in $\lambda$ and~$t$,
$D^{(1)}$ is a bivariate polynomial of total degree~$n$, monic of degree $n$ in~$\lambda$,
and of degree at most $n$ in~$t$ (with equality if and only if $\det A\neq0$).

As in \cref{subsec:prelimin}, we use as a black box a fixed algorithm which multiplies two $n\times n$
matrices over~$\R$ using~$O(n^{\omega})$ ring operations.
\Cref{alg:charpoly} only uses ring additions, subtractions and multiplications, with no division and no
zero test, and its control depends only on~$n$.

\begin{theorem}\label{thm:charpoly}
Let $\R$ be a commutative ring with unit and $A\in\R^{n\times n}$. If $\omega>2$, \cref{alg:charpoly} computes the characteristic polynomial  of $A$
using $O(n^{\omega})$ operations in~$\R$.
\end{theorem}
\begin{proof}
Up to the padding of \cref{subsec:overview}, which multiplies the cost by at most a constant factor,
we may assume that $n$ is a power of two. With $A'=\diag(A,I_{n'-n})$, $\det(\lambda I-A)$ is the exact quotient of $\det(\lambda I-A')$ by
$(\lambda-1)^{n'-n}$. The divisor being monic, the division to recover $\det(\lambda I-A)$ uses only ring operations, and costs $O(n^2)$.

\Cref{step:init} gives the exact terms $H^{(n)}_0$ and $H^{(n)}_1$.
By \cref{lem:transition}, each iteration of \cref{step:loop} gives the exact terms $H^{(m/2)}_0,\ldots,H^{(m/2)}_{4d-1}$,
which are the input of the next iteration.  At $m=2$ these are $H^{(1)}_0,\ldots,H^{(1)}_{2n-1}$, 
which then lead to $D^{(1)}$ modulo $t^{n+1}$ at level $m=1$, using \cref{lem:lifting} with $\sigma=n+1$. 

By \cref{lem:detD}, $D^{(1)}(\lambda,t)=\det(\lambda I-L(t))$. Since $L$ has degree one in~$t$,
this polynomial has degree at most $n$ in~$t$, and it is thus computed exactly.
Its constant coefficient is $D^{(1)}_0=(-1)^n\det L=(-1)^n\det(I-tA)\det M=-\det(I-tA)$,
since $\det M=(-1)^{n-1}$. It follows that $c(t)=\det(I-tA)$ and $\lambda^nc(1/\lambda)=\det(\lambda I-A)$.

For the cost, we take $\mathsf M(h)=O(h\log h\log\log h)$ \cite{CaKa91}
(more generally, any $\mathsf M(h)=O(h^{\omega/2-\varepsilon})$ with $\varepsilon>0$ would do).
By \cref{lem:transition}, with $m=n/d$ and $d=2^i$, the iterations of \cref{step:loop} cost
\[
O\Bigl(\sum_{i=0}^{\log n-1}\Bigl(\frac{n}{2^i}\Bigr)^{\omega}\mathsf M(4^i)\,(1+i)\Bigr)
=O\Bigl(n^{\omega}\sum_{i=0}^{\log n-1}2^{-(\omega-2)i}\,(1+i)^2\log(2+i)\Bigr).
\]
For $\omega>2$ the sum is bounded, which gives $O(n^{\omega})$.
\Cref{step:last} costs $O(\mathsf M(n^2)\log n)$ by \cref{lem:lifting}, which is $O(n^{\omega})$ for $\omega>2$,
and the final reversal is free.
\end{proof}

In particular, $\det A$ is $(-1)^n$ times the constant term of the characteristic polynomial,
and 
\Cref{alg:charpoly} also computes the bivariate polynomial $D^{(1)}=\det(\lambda I-L(t))\in\R[\lambda,t]$.
If $\omega=2$, the analysis for \cref{thm:charpoly} gives $O(n^2\log^3n\log\log n)$ operations in~$\R$,
the levels having then comparable costs.

Note also that the cyclic permutation $M$ could be replaced by any companion matrix whose constant coefficient is a unit of~$\R$ with a known inverse
(the result is then divided by the known unit $\det M$).
In the same spirit as the application of~\cite{KaVi05} to integer matrices, this leads to the characteristic polynomial
of a polynomial matrix with cyclic constant term, over a field, which we leave for future work.

\bibliographystyle{abbrvnat}

{\footnotesize

\bibliography{biblio}

}

\end{document}